\documentclass[12pt,twoside]{article}

\usepackage{graphicx}
\usepackage{epsfig}
\usepackage{url}
\usepackage{psfrag}
\usepackage{times}
\usepackage{amsfonts,amssymb,amsbsy}
\usepackage{amsmath}
\usepackage{latexsym}
\usepackage{amsfonts,amsbsy}
\usepackage{latexsym}

\begin{document}

\def\e{\begin{equation}} 
\def\f{\end{equation}} 
\def\ea{\begin{eqnarray}} 
\def\fa{\end{eqnarray}} 

\def\##1{{\mbox{\textbf{#1}}}}
\def\%#1{{\mbox{\boldmath $#1$}}}
\def\=#1{{\overline{\overline{\mathsf #1}}}}
\def\RR{\mbox{\boldmath $\R$}}
\def\nn#1{{\sf #1}}
\def\SE{{\mathbb E}}
\def\SF{{\mathbb F}}

\def\*{^{\displaystyle*}}
\def\xx{\displaystyle{{}^\times}\llap{${}_\times$}}
\def\.{\cdot}
\def\x{\times}
\def\oo{\infty}

\def\D{\nabla}
\def\d{\partial}

\def\ra{\rightarrow}
\def\lra{\leftrightarrow}
\def\Ra{\Rightarrow}
\def\le{\left(}
\def\ri{\right)}
\def\l#1{\label{eq:#1}}
\def\r#1{(\ref{eq:#1})}
\def\am{\left(\begin{array}{c}}
\def\amm{\left(\begin{array}{cc}}
\def\ammm{\left(\begin{array}{ccc}}
\def\ammmm{\left(\begin{array}{cccc}}
\def\a{\end{array}\right)}

\def\I{\int\limits}
\def\OI{\oint\limits}

\def\A{\alpha}
\def\B{\beta}
\def\de{\delta}
\def\De{\Delta}
\def\E{\epsilon}
\def\g{\gamma}
\def\G{\Gamma}
\def\h{\eta}
\def\K{\kappa}
\def\la{\lambda}
\def\La{\Lambda}
\def\M{\mu}
\def\o{\omega}
\def\Om{\Omega}
\def\R{\rho}
\def\s{\sigma}
\def\t{\tau}
\def\z{\zeta}
\def\X{\chi}
\def\TH{\theta}
\def\Th{\Theta}
\def\VF{\varphi}
\def\VR{\varrho}
\def\VT{\vartheta}
\def\ve{\%\varepsilon}

\def\tr{{\rm tr }}
\def\spm{{\rm spm}}
\def\det{{\rm det}}
\def\Det{{\rm Det}}
\def\sgn{{\rm sgn}}

\def\W{\wedge}
\def\WW{\displaystyle{{}^\wedge}\llap{${}_\wedge$}}
\def\Adj{{\rm Adj\mit}}
\def\ua{\uparrow}
\def\da{\downarrow}
\def\uda{\updownarrow}

\def\J{\rfloor}
\def\L{\lfloor}
\def\JJ{\rfloor\rfloor}
\def\LL{\lfloor\lfloor}


\title{Generalized Soft-and-Hard/DB Boundary}
\author{I.V. Lindell and A. Sihvola{${}^1$}} 
\date{Department of Radio Science and Engineering,\\ Aalto University, Espoo, Finland\\ 
{\tt ismo.lindell@aalto.fi}\\\vspace{-1pt}{\tt ari.sihvola@aalto.fi}}
\pagestyle{myheadings}

\pagestyle{myheadings}

\textwidth=16cm
\hoffset=-1.5cm
\voffset=-2cm
\textheight=24cm
\parindent=0pt
\parskip=\medskipamount

\maketitle

\begin{abstract}
A novel class of boundary conditions is introduced as a generalization of the previously defined class of soft-and-hard/DB (SHDB) boundary conditions. It is shown that the conditions for the generalized soft-and-hard/DB (GSHDB) boundary arise most naturally in a simple and straightforward manner by applying four-dimensional differential-form and dyadic formalism. At a given boundary surface, the GSHDB conditions are governed by two one-forms. In terms of Gibbsian 3D vector and dyadic algebra the GSHDB conditions are defined in terms of two vectors tangential to the boundary surface and two scalars. Considering plane-wave reflection from the GSHDB boundary, for two eigenpolarizations, the GSHDB boundary can be replaced by the PEC or PMC boundary. Special attention is paid to the problem of plane waves matched to the GSHDB boundary, defined by a 2D dispersion equation for the wave vector, making the reflection dyadic indeterminate. Examples of dispersion curves for various chosen parameters of the GSHDB boundary are given. Conditions for a possible medium whose interface acts as a GSHDB boundary are discussed. 
\end{abstract}

\section{Introduction}

Recent progress in metamaterials and metasurfaces has created interest in theoretical analysis of general classes of electromagnetic media and boundaries, defined by medium equations and boundary conditions \cite{Eleftheriades18} -- \cite{Wen15}. In the present study, the previously defined SHDB boundary \cite{SHDB}, \cite{Surf} which is a generalization of both the soft-and-hard (SH) boundary \cite{SHS1} and the DB boundary \cite{DB}, is generalized one step further. 

It has turned out that basic boundary conditions can be expressed in simple form by applying four-dimensional formalism \cite{Deschamps,MDEM} according to which the fields are represented by two electromagnetic two-forms as
\e \%\Phi = \#B + \#E\W\ve_4,\ \ \ \ \%\Psi= \#D - \#H\W\ve_4. \f
Here $\#B,\#D$ are spatial two-forms and $\#E,\#H$ are spatial one-forms. $\ve_4=c\#d t$ represents the temporal one-form in a basis $\{\ve_i\}$, while $\ve_1,\ve_2$ and $\ve_3$ are spatial one-forms. For a more detailed exposition of definitions and properties related to this formalism and corresponding 3D Gibbsian vector representations, one should consult \cite{MDEM} or \cite{Difform}. Denoting a planar boundary by $\ve_3|\#x=x_3=0$, where $\ve_3$ is a constant spatial one-form and $\#x=\sum_1^4\#e_i x_i$ is the four-vector, we can define a set of different basic boundary conditions as follows. 

\begin{itemize}
\item Perfect electric conductor (PEC) boundary is defined by the 4D condition
\e \ve_3\W\%\Phi=0\ \ \ \ \Ra\ \ \ \ \ve_3\W\#B=0,\ \ \ve_3\W\#E=0. \l{PEC} \f
The corresponding conditions for the Gibbsian vector fields obtained as \cite{MDEM}
\e \#B_g = \#e_{123}\L\#B,\ \ \ \#E_g= \=G_s|\#E, \f
with the spatial metric dyadic defined by
\e \=G_s = \#e_1\#e_1+ \#e_2\#e_2 + \#e_3\#e_3,\f
become
\e \#e_3\.\#B_g=0,\ \ \ \ \#e_3\x\#E_g=0. \l{PECg}\f
Boundary conditions of the form \r{PEC} are basic in the sense that they involve no parameters.
\item Perfect magnetic conductor (PMC) boundary is defined similarly as 
\e \ve_3\W\%\Psi = 0\ \ \ \ \Ra\ \ \ \ \ve_3\W\#D=0,\ \ \ve_3\W\#H=0, \l{PMC} \f
and, for the Gibbsian vector fields,
\e \#D_g= \#e_{123}\L\#E,\ \ \ \ \#H_g=\=G_s|\#H, \f
as
\e \#e_3\.\#D_g=0,\ \ \ \ \#e_3\x\#H_g=0. \l{PMCg}\f
There is no parameter in \r{PMC}, either.
\item Perfect electromagnetic conductor (PEMC) boundary \cite{PEMC}, defined by
\e \ve_3\W(\%\Psi-M\%\Phi)=0, \l{PEMC}\f
is a generalization of PMC ($M=0$) and PEC ($1/M=0$) boundaries. \r{PEMC} can be  split in its spatial and temporal parts as
\e \ve_3\W(\#D-M\#B)=0,\ \ \ \ \ve_3\W(\#H+M\#E)=0,\f
which correspond to the Gibbsian conditions,
\e \#e_3\.(\#D_g-M\#B_g)=0,\ \ \ \ \#e_3\x(\#H_g+M\#E_g)=0. \f
The PEMC boundary involves one scalar parameter $M$, the PEMC admittance.
\item The DB boundary is defined by the two conditions \cite{MDEM},
\e \ve_3\W\ve_4\W\%\Phi=0\ \ \Ra\ \ \ve_3\W\#B=0,\l{DBB}\f
\e \ve_3\W\ve_4\W\%\Psi=0\ \ \Ra\ \ \ve_3\W\#D=0, \l{DBD}\f
which correspond to the Gibbsian conditions \cite{DB}
\e \#e_3\.\#B_g=0,\ \ \ \ \#e_3\.\#D_g=0. \f
The conditions \r{DBB}, \r{DBD} of the DB boundary do not involve any parameters.
\item The soft-and-hard (SH) boundary, originally introduced by P.-S. Kildal \cite{SHS1},  is defined by the 4D conditions 
\e \ve_3\W\%\A_s\W\%\Phi=0\ \ \Ra\ \ \ve_3\W\%\A_s\W\#E= 0,\f
\e \ve_3\W\%\A_s\W\%\Psi=0\ \ \Ra\ \ \ve_3\W\%\A_s\W\#H=0, \f
where $\%\A_s$ is a spatial one-form satisfying $\ve_3\W\%\A_s\not=0$. Defining a suitable basis, we can set $\%\A_s=\ve_1$, whence the Gibbsian conditions become 
\e \#e_3\x\#e_1\.\#E_g= \#e_2\.\#E_g=0,\ \ \ \ \#e_3\x\#e_1\.\#H_g = \#e_2\.\#H_g=0. \f
The SH boundary conditions involves one parameter defining the vector $\#e_2$ orthogonal to $\#e_3$.
\item As a generalization of both SH and DB boundary conditions we can set
\e \ve_3\W\%\A\W\%\Phi=0,\ \ \ \ \ve_3\W\%\A\W\%\Psi=0, \l{SHDB}\f
where the one-form $\%\A$ may have both a spatial and a temporal component. Such conditions have been dubbed as those of the soft-and-hard/DB (SHDB) boundary \cite{SHDB}. Without losing generality, we can set
\e \%\A = \A_1\ve_1 + \A_4\ve_4, \f
whence the conditions \r{SHDB} can be expanded as
\e \A_4\ve_3\W\#B + \A_1\ve_3\W\ve_1\W\#E =0,\f
\e \A_4\ve_3\W\#D - \A_1\ve_3\W\ve_1\W\#H =0. \f
The corresponding Gibbsian conditions are now
\e \A_4\#e_3\.\#B_g + \A_1\#e_2\.\#E_g =0,\f
\e \A_4\#e_3\.\#D_g - \A_1\#e_2\.\#H_g =0. \f
In this case there are two free parameters ($\A_1/\A_4$, $\#e_2$), defining the SHDB boundary at the surface $\ve_3|\#x=0$ or $\#e_3\.\#r=0$. 
 
\end{itemize}

Properties of plane waves reflecting from the SHDB boundary have been studied previously  \cite{SHDB}. Most notably, it has been shown that when a given incident plane wave is split in two plane waves, with polarizations depending on the direction of incidence and the parameters of the boundary in a certain manner, one of them is reflected as from the PEC boundary, and the other one, as from the PMC boundary. The same property is also valid for the SH and DB special cases. As another important property, the SHDB boundary was shown to be self dual, i.e., invariant in a duality transformation changing electric and magnetic quantities to one another. This property is not shared by the PEC and PMC boundaries or the PEMC boundary. In fact, PEC and PMC boundaries are transformed to one another, while a PEMC boundary is transformed to another PEMC boundary. Realizations of various boundary conditions as metasurfaces have been reported in \cite{Caloz} -- \cite{Zaluski2014}, \cite{Frezza}, and some applications have been pointed out in \cite{Kong08} and 
\cite{Yaghjian}. 

It is the purpose of the present paper to study a natural generalization of the SHDB boundary and study its properties in plane-wave reflection.

\section{Generalized soft-and-hard/DB (GSHDB) boundary}

As an obvious generalization of the SHDB boundary conditions \r{SHDB} we may consider
\e \ve_3\W\%\A\W\%\Phi= 0,\ \ \ \ve_3\W\%\B\W\%\Psi=0, \f
where $\%\A$ and $\B$ are two one-forms. Expanded in terms of spatial and temporal parts, 
\e \%\A = \%\A_s + \A_4\ve_4,\ \ \ \ \%\A_s=\A_1\ve_1+ \A_2\ve_2, \f
\e \%\B = \%\B_s + \B_4\ve_4,\ \ \ \ \%\B_s=\B_1\ve_1+\B_2\ve_2, \f
we have
\ea \A_4+\ve_3\W\#B + \ve_3\W\%\A_s\W\#E&=& 0 \\
\B_4+\ve_3\W\#D - \ve_3\W\%\B_s\W\#H&=& 0 \fa
The corresponding Gibbsian conditions become
\ea \A_4\#e_3\.\#B_g + \%\A_g\.\#E_g&=& 0, \l{Bg}\\
\B_4\#e_3\.\#D_g - \%\B_g\.\#H_g&=& 0 , \l{Dg}\fa
in terms of the Gibbsian vectors
\e \%\A_g =\A_1\#e_1+ \A_2\#e_2,\ \ \ \ \%\B_g=\B_1\#e_1+ \B_2\#e_2. \f
The unit vectors $\#e_1,\#e_2,\#e_3$ are assumed to make an orthonormal basis. Let us call the conditions \r{Bg}, \r{Dg} as those of the generalized soft-and-hard/DB (GSHDB) boundary. 

The GSHDB boundary conditions have a number of special cases.
\begin{itemize}
\item The SHDB boundary conditions are obtained when the parameters satisfy the relations
\e A\A_4+B\B_4=0,\ \ \ \ \ A\%\A_g+B\%\B_g=0, \f
for some scalars $A,B$.
\item Generalized soft-and-hard (GSH) boundary conditions are obtained for $\A_4=\B_4=0$ as
\e \%\A_g\.\#E_g=0,\ \ \ \ \%\B_g\.\#H_g= 0 , \l{GSH}\f
with $\#e_3\.\%\A_g=\#e_3\.\%\B_g=0$. \r{GSH} reduces to the SH conditions when $\%\A_g$ and $\%\B_g$ are multiples of the same vector. The GSH boundary has been previously considered in \cite{GSH,RGSH}.
\item The DB boundary conditions are obtained for $\%\A_g=\%\B_g=0$.
\item As two simple special cases of the GSHDB boundary, neither of which falls into the class of SHDB boundaries, we may consider those defined by the conditions
\e \#e_3\.\#B_g=0,\ \ \ \ \%\B_g\.\#H_g=0, \f
and
\e \#e_3\.\#D_g=0,\ \ \ \ \%\A_g\.\#E_g=0. \f
The two vectors are assumed to satisfy $ \#e_3\.\%\B_g=\#e_3\.\%\A_g=0$.
\end{itemize}

\section{Plane-wave reflection from GSHDB boundary}

Above, the 4D formalism has been applied mainly to demonstrate the simple algebraic definition of the GSHDB boundary conditions. Let us now proceed with the Gibbsian 3D formalism by assuming vector fields everywhere and drop the subscript $()_g$. The conditions \r{Bg}, \r{Dg} can now be written as
\ea \A_4\#e_3\.\#B + \%\A_t\.\#E_t &=& 0, \l{BAE} \\
\B_4\#e_3\.\#D -\%\B_t\.\#H_t&=&0, \l{DBH} \fa
where vectors transverse to $\#e_3$ have been emphasized by the subscript $()_t$. Assuming that the medium in the half space $\#e_3\.\#r>0$ above the planar GSHDB boundary is isotropic with parameters $\E_o,\M_o$, the conditions \r{BAE}, \r{DBH} can be written as
\ea \A_o\#e_3\.\h_o\#H + \%\A_t\.\#E_t&=& 0, \l{Bg1}\\
\B_o\#e_3\.\#E - \%\B_t\.\h_o\#H_t&=& 0 , \l{Dg1}\fa
with
\e \A_o = \A_4\sqrt{\M_o\E_o},\ \ \ \ \B_o = \B_4\sqrt{\M_o\E_o},\ \ \ \ \h_o=\sqrt{\M_o/\E_o}. \l{AoBo}\f

\subsection{Plane-wave conditions}

Let us consider a time-harmonic plane wave of the form $\#E^i(\#r,t)=\#E^i\exp(j(\o t-\#k^i\.\#r))$ incident to the GSHDB boundary with the wave vector $\#k^i=-k_3\#e_3+ \#k_t$. The reflected wave $\#E^r(\#r,t)=\#E^r\exp(j(\o t-\#k^r\.\#r))$ depends on the wave vector $\#k^r=k_3\#e_3 + \#k_t$. The two wave vectors satisfy
\e \#k^i\.\#k^i=\#k^r\.\#k^r = k_3^2+\#k_t\.\#k_t=k_o^2,\ \ \ \ k_o=\o\sqrt{\M_o\E_o}. \f 
Conditions for the plane-wave fields
\ea \#k^i\x\#E^i &=& \o\#B^i= \o\M_o\#H^i,\\
 \#k^r\x\#E^r &=& \o\#B^r= \o\M_o\#H^r, \l{kE}\fa
\ea \#k^i\x\#H^i &=& -\o\#D^i= -\o\E_o\#E^i,\\
 \#k^r\x\#H^r &=& -\o\#D^r=-\o\E_o\#E^r, \l{kH}\fa
are obtained from the Maxwell equations. 

Relations between the fields $\#E_t^i$ and $\#H_t^i$ on one hand, and $\#E_t^r$ and $\#H_t^r$ on the other hand, can be found by applying the orthogonality conditions $\#k^i\.\#E^i = \#k^r\.\#E^r=\#k^i\.\#H^i = \#k^r\.\#H^r=0$, whence the field vectors can be expressed in terms of their tangential components as
\ea k_3\#E^i &=& (\#e_3\#k_t + k_3\=I_t)\.\#E_t^i, \l{k3Ei}\\
k_3\#E^r &=& -(\#e_3\#k_t-k_3\=I_t)\.\#E_t^r,\\
 k_3\#H^i &=& (\#e_3\#k_t+k_3\=I_t)\. \#H_t^i, \\
k_3\#H^r &=& -(\#e_3\#k_t-k_3\=I_t)\. \#H_t^r, \l{k3Hr}\fa
where
\e \=I_t = \#e_1\#e_1+ \#e_2\#e_2 \f
is the tangential unit dyadic. The transverse components can be found to satisfy the relations
\ea \h_o\#H^i_t  &=& -\=J_t\.\#E_t^i ,\l{kEit}\\
  \h_o\#H^r_t  &=&\=J_t\.\#E_t^r , \l{kErt}\\
  \#E^i_t  &=&\=J_t\.\h_o\#H_t^i ,\l{kHit} \\
\#E^r_t   &=& -\=J_t\.\h_o\#H_t^r , \l{kHrt}\fa
in terms of the two-dimensional dyadic
\e \=J_t = \frac{1}{k_ok_3}((\#e_3\x\#k_t)\#k_t+k_3^2\#e_3\x\=I_t), \l{Jt}\f
excluding the case of incidence parallel to the boundary, $k_3=0$. From \r{kEit} -- \r{kHrt} we obtain
\e \=J{}_t^2 = -\=I_t, \ \ \ \ \ \=J{}_t^{-1}=-\=J_t,\l{J2}\f
which can also be verified from the expression \r{Jt}. Thus, the dyadic $\=J_t$ resembles the imaginary unit. It also satisfies the properties
\e \#e_3\#e_3\xx\=J_t = -\=J{}_t^T,\ \ \ \ \=J{}_t^{(2)}=\frac{1}{2}\=J_t\xx\=J_t= \#e_3\#e_3, \l{eeJ}\f
\e \tr\=J_t =0,\ \ \ \ \ \det\=J_t = \tr\=J{}_t^{(2)}= 1, \f
which can be easily verified.

\subsection{Reflection dyadic}

Applying \r{k3Ei} -- \r{k3Hr}, the GSHDB boundary conditions \r{Bg1} and \r{Dg1} can be reduced to the respective conditions
\e \#a_t\.(\#E_t^i+\#E_t^r) =0,\l{aEir} \f
\e \#b_t\.(\#H_t^i+\#H_t^r)=0, \l{bHir}\f
when the two vectors tangential to the boundary are defined by
\ea \#a_t &=& \A_o\#e_3\x\#k_t + k_o\%\A_t, \l{a}\\
\#b_t &=& \B_o\#e_3\x\#k_t +k_o\%\B_t, \l{b}\fa
and $\A_o,\B_o$ by \r{AoBo}. One should note that, for the special case of the SHDB boundary, the two vectors are the same, $\#a_t=\#b_t$.

Applying \r{kEit} -- \r{kHrt}, the GSHDB boundary conditions \r{aEir} and \r{bHir} for fields in an isotropic medium can be further written as
\ea \#a_t\.(\#E_t^i+\#E_t^r) &=& \h_o\#d_t\.(\#H_t^i-\#H_t^r) =0, \l{atdt}\\
\h_o\#b_t\.(\#H_t^i+\#H_t^r) &=& \#c_t\.(\#E_t^i-\#E_t^r)=0, \l{btct}\fa
where the two additional vectors are defined by
\ea \#c_t &=& \#b_t\.\=J_t \nonumber\\
&=& \frac{1}{k_ok_3}((\#e_3\.\#k_t\x\#b_t)\#k_t-k_3^2(\#e_3\x\#b_t)), \l{c}\\
\#d_t &=& \#a_t\.\=J_t \nonumber\\
&=& \frac{1}{k_ok_3}((\#e_3\.\#k_t\x\#a_t)\#k_t-k_3^2(\#e_3\x\#a_t)). \l{d} \fa
For the special case of the SHDB boundary we have $\#c_t=\#d_t$. 

The four vectors satisfy
\e \#a_t\x\#c_t = \#b_t\x\#d_t = \#e_3 \De, \l{acbd}\f
with
\ea \De &=& \#e_3\.\#a_t\x\#c_t  = -\#a_t\.(\#e_3\x\=J{}_t^T)\.\#b_t \nonumber\\
&=& -\frac{1}{k_ok_3}\#a_t\.(k_3^2\=I_t +(\#e_3\x\#k_t)(\#e_3\x\#k_t))\.\#b_t \nonumber\\
&=& \frac{k_o}{k_3}((\#k_t\.\%\A_t)(\#k_t\.\%\B_t) \nonumber\\
&-&(\A_o\#e_3\x\#k_t + k_o\%\A_t)\.(\B_o\#e_3\x\#k_t +k_o\%\B_t)). \l{De}\fa

Assuming $\De\not=0$, in which case the vectors $\#a_t$ and $\#c_t$ on one hand, and $\#b_t$ and $\#d_t$ on the other hand, are linearly independent, the relations between reflected and incident  fields can be found by applying  \r{acbd}, \r{atdt} and \r{btct} in
\ea \De\#e_3\x\#E_t^r &=& (\#a_t\x\#c_t)\x\#E^r = (\#c_t\#a_t-\#a_t\#c_t)\.\#E^r \nonumber\\
&=&- (\#c_t\#a_t+\#a_t\#c_t)\.\#E^i, \fa
\ea \De\#e_3\x\#H_t^r &=& (\#b_t\x\#d_t)\x\#H^r = (\#d_t\#b_t-\#b_t\#d_t)\.\#H^r \nonumber\\
&=& -(\#d_t\#b_t+ \#b_t\#d_t)\.\#H^i. \fa
The reflected fields can be solved as
\e \#E_t^r = \=R_E\.\#E_t^i,\ \ \ \ \ \#H_t^r=\=R_H\.\#H_t^i, \f
in terms of two 2D reflection dyadics defined by
\ea \=R_E &=& \frac{1}{\De}\#e_3\x (\#c_t\#a_t+\#a_t\#c_t), \l{RE}\\
\=R_H &=& \frac{1}{\De}\#e_3\x(\#d_t\#b_t+\#b_t\#d_t). \l{RH}\fa 
The case $\De=0$ will be considered in Section IV. The two reflection dyadics satisfy
\e \tr\=R_E=\tr\=R_H=0, \f
\e \=R{}_E^{(2)}=  -\#e_3\#e_3, \ \ \ \  \det\=R_E = \tr\=R{}_E^{(2)} = -1, \f
\e \=R{}_H^{(2)}= -\#e_3\#e_3, \ \ \ \  \det\=R_H = \tr\=R{}_H^{(2)}  = -1, \f
as can be easily verified. They coincide for the special case of the SHDB boundary, $\=R_E=\=R_H$. 

From \r{acbd} and $\De\not=0$ it follows that the vectors $\#a_t,\#c_t,\#e_3$ and $\#b_t,\#d_t,\#e_3$ make two vector bases whose respective reciprocal basis vectors can be constructed as \cite{Methods}
\ea \#a_t' = \frac{1}{\De}\#c_t\x\#e_3,\ \ \ \#c_t' = \frac{1}{\De}\#e_3\x\#a_t,\ \ \ \#e_3'=\#e_3, \l{ac'}\\
 \#b_t' = \frac{1}{\De}\#d_t\x\#e_3,\ \ \ \#d_t' = \frac{1}{\De}\#e_3\x\#b_t,\ \ \ \#e_3'=\#e_3. \l{bd'}\fa
They satisfy the conditions
\e \#a_t\.\#a_t'=\#c_t\.\#c_t' = \#b_t\.\#b_t'=\#d_t\.\#d_t'=1, \f
\e \#a_t\.\#c_t'=\#c_t\.\#a_t' = \#b_t\.\#d_t'=\#d_t\.\#c_t'=0, \f
whence the tangential unit dyadic can be expressed as
\e \=I_t = \#a'_t\#a_t + \#c'_t\#c_t = \#b'_t\#b_t+ \#d'_t\#d_t. \f
From \r{eeJ}, \r{c} and \r{d}, we obtain the relations
\e \=J_t\.\#a_t'=-\#d_t',\ \ \ \ \=J_t\.\#b_t'=-\#c_t', \l{Jab}\f
whence the reflection dyadics \r{RE} and \r{RH} can be written as
\e \=R_E = -\#a_t'\#a_t+ \#c_t'\#c_t,\ \ \ \ \=R_H=-\#b_t'\#b_t + \#d_t'\#d_t. \l{REH}\f 
They satisfy
\e \=R{}_E^2=\#a_t'\#a_t+ \#c_t'\#c_t=\=I_t,\ \ \ \ \ \=R{}_H^2=\#b_t'\#b_t + \#d_t'\#d_t=\=I_t. \f

Relations between the two reflection dyadics can be found through the chain
\e \h_o\#H_t^r = \=J_t\.\#E_t^r =\=J_t\.\=R_E\.\#E_t^i = \=J_t\.\=R_E\.\=J_t\.\h_o\#H_t^i \f
as
\e \=R_H = \=J_t\.\=R_E\.\=J_t,\ \ \ \ \=R_E = \=J_t\.\=R_H\.\=J_t. \f
They can be verified by inserting \r{REH} and applying \r{c}, \r{d} and \r{Jab}.

\subsection{Eigenfields}

Because from \r{REH} we have
\e \=R_E\.\#a_t' = -\#a_t',\ \ \ \=R_E\.\#c_t' = \#c_t', \f 
and
\e \=R_H\.\#b_t' = -\#b_t',\ \ \ \=R_H\.\#d_t' = \#d_t', \f 
the eigenvalues of the two eigenproblems involving tangential fields  
\e \#E_t^r= \=R_E\.\#E_t^i = \la_E\#E_t^i, \ \ \ \ \#H_t^r=\=R_H\.\#H_t^i= \la_H\#H_t^i \f
are $+1$ and $-1$. Let us label the two eigenwaves as the e-wave and the m-wave. Applying \r{kEit} -- \r{kHrt} we obtain $\la_E=-\la_H$, whence we may set
\e \la_E=-1,\ \ \ \la_H=+1,\ \ \ \ {\rm (e-wave)}, \f
\e \la_E=+1,\ \ \ \la_H=-1,\ \ \ \ {\rm (m-wave)}. \f
The eigenfields obey the relations
\ea \#E_{et}^i=-\#E_{et}^r&=& E_e^i\#a_t',\\
 \h_o\#H_{et}^i=\h_o\#H_{et}^r &=& -\=J_t\.\#E_{et}^i = E_e^i\#d_t', \\ 
\#E_{mt}^i=\#E_{mt}^r&=& E_m^i\#c_t',\\
 \h_o\#H_{mt}^i=-\h_o\#H_{mt}^r &=& -\=J_t\.\#E_{mt}^i = -E_m^i\#b_t', \fa 
where we have again applied the plane-wave rules \r{kEit} -- \r{kHrt} and \r{Jab}.

The problem of finding the reflection of a given incident plane wave can now be solved by expanding the incident plane wave in its e-wave and m-wave components,
\ea \#E_t^i &=& (\#a_t'\#a_t+ \#c_t'\#c_t)\.\#E_t^i= E_e^i\#a_t' + E_m^i\#c_t',\\
\#H_t^i &=& (\#b_t'\#b_t + \#d_t'\#d_t)\.\#H_t^i= H_m^i\#b_t' + H_e^i\#d_t'. \fa
The reflected wave becomes
\ea \#E_t^r &=& (-\#a_t'\#a_t+ \#c_t'\#c_t)\.\#E_t^i= E_e^r\#a_t' + E_m^r\#c_t',\\
\#H_t^r &=& (-\#b_t'\#b_t + \#d_t'\#d_t)\.\#H_t^i= H_m^r\#b_t' + H_e^r\#d_t', \fa
with the field magnitudes obtained from
\e E_e^i=-E_e^r=\#a_t\.\#E_t^i,\ \ \ \ \ \ E_m^i=E_c^r=\#c_t\.\#E_t^i, \f
\e H_m^i=-H_m^r=\#b_t\.\#H_t^i,\ \ \ \ \ \ H_e^i=H_e^r=\#d_t\.\#H_t^i . \f

The total fields at the boundary $\#e_3\.\#r=0$ are reduced to
\ea \#E_t^i+\#E_t^r &=& 2E_m^i\#c',\\
 \h_o(\#H_t^i+\#H_t^r) &=& 2E_e^i\#d_t', \\
\o\#e_3\.(\#B^i+\#B^r) &=&\#e_3\.\#k_t\x(\#E^i+\#E^r) \nonumber\\
&=& 2E_m^i\#e_3\.\#k_t\x\#c',\\
-\o\#e_3\.(\#D^i+\#D^r) &=&\#e_3\.\#k_t\x(\#H^i+\#H^r) \nonumber\\
&=& 2E_e^i\#e_3\.\#k_t\x\#d_t'. \fa
Because we have
\e \#e_3\x(\#E_e^i+\#E_e^r)=0,\ \ \ \ \#e_3\.(\#B_e^i+\#B_e^r)=0, \f
the e-wave satisfies the PEC conditions \r{PECg} at the GSHDB boundary. Similarly, from
\e \#e_3\x(\#H_m^i+\#H_m^r)=0,\ \ \ \ \#e_3\.(\#D_m^i+\#D_m^r)=0, \f
the m-wave satisfies the PMC conditions \r{PMCg} at the GSHDB boundary. Thus, the property known for the SHDB boundary \cite{SHDB} and its special cases, the SH boundary and the DB boundary, is preserved in the generalization to the GSHDB boundary. Actually, the e-wave satisfies the GSHDB condition \r{btct} everywhere, while the condition \r{atdt} required at the boundary equals the PEC condition for the e-wave. Similarly, the m-wave satisfies the GSHDB condition \r{atdt} everywhere while the condition \r{btct} equals the PMC condition at the boundary.

\section{Plane waves matched to GSHDB boundary}

In the previous analysis it was assumed that the scalar \r{De} does not vanish. Let us now study the converse case, $\De=0$, which may happen for some special values of the wave vector $\#k^i$. Assuming $k_3\not=0$, the condition can be written as
$$\#k_t\.(\%\A_t\%\B_t -\A_o\B_o\=I_t)\.\#k_t +$$
\e +\#k_t\.(k_o\#e_3\x(\A_o\%\B_t+ \B_o\%\A_t)) -k_o^2\%\A_t\.\%\B_t = 0, \l{De0}\f
which is a quadratic equation for $\#k_t$. The corresponding vectors $\#k^i$ and $\#k^r$ are uniquely determined by $\#k_t$. 

Because from \r{acbd} we have
\e \#a_t\x\#c_t = \#b_t\x\#d_t = 0, \l{atct0}\f
the vectors $\#a_t$ and $\#c_t$ on one hand, and $\#b_t$ and $\#d_t$ on the other hand, are multiples of one another. Thus, the steps leading to the reflection dyadics \r{RE}, \r{RH} are not valid for such wave vectors $\#k^i$. 
From \r{atdt} and \r{btct} we obtain
\e \#a_t\.\#E^i=\#a_t\.\#E^r=0,\ \ \ \ \#d_t\.\#H^i=\#d_t\.\#H^r=0. \f 
whence the field vectors can be expressed as
\ea \#E^i_t &=& E^i\#e_3\x\#a_t,\\
 \#E^r_t&=&E^r\#e_3\x\#a_t, \\
\h_o\#H^i_t &=& -\=J_t\.\#E_t^i =  -E^i\=J_t\.(\#e_3\x\#a_t) \nonumber\\
&=& E^i\#e_3\x\#d_t, \\
\h_o\#H_t^r&=& \=J_t\.\#E_t^r =  E^r\=J_t\.(\#e_3\x\#a_t) \nonumber\\
&=& -E^r\#e_3\x\#d_t. \fa
Here we have applied \r{eeJ}. It is remarkable that there is no relation between the magnitudes of the incident and reflected fields. Because each of them satisfies the GSHDB conditions, they may exist independently and can be called plane waves matched to the GSHDB boundary. In particular, if $k_3$ has a complex value and $\#E^i$ decays exponentially away from the boundary, it is known as a surface wave, while $\#E^r$ is known as a leaky wave due to its exponential growth. 

The condition \r{De0} is an equation restricting the vector $\#k_t$ of a plane wave match\-ed to a given GSHDB boundary defined by the parameters $\A_o,\%\A_t,\B_o,\%\B_t$. Actually, \r{De0} can be considered as the 2D counterpart of the 3D dispersion equation $D(\#k)=0$ restricting the wave vector $\#k$ of a plane wave in an electromagnetic medium and we may call \r{De0}  the dispersion equation for the plane wave matched to a boundary. Expanding 
\e \#k_t=\#u_t k_t= k_t(\#e_1\cos\VF + \#e_2\sin\VF)\, \f
\e \%\A_t=\#e_1\A_1+\#e_2\A_2,\ \ \ \%\B_t=\#e_1\B_1+\#e_2\B_2, \f
the two solutions of the dispersion equation \r{De0} can be expressed for a given tangential unit vector $\#u_t$ as
\e k_t(\#u_t) = k_o(A \pm \sqrt{A^2+B}), \l{ktu}\f
with
\e A= \frac{(\#e_3\x\#u_t)\.(\A_o\%\B_t+\B_o\%\A_t)}{2((\#u_t\.\%\A_t)(\#u_t\.\%\B_t)-\A_o\B_o)} \f
$$ = \frac{(\A_o\B_2+\B_o\A_2)\cos\VF-(\A_o\B_1+\B_o\A_1)\sin\VF}{2((\A_1\cos\VF+\A_2\sin\VF)(\B_1\cos\VF+\B_2\sin\VF)-\A_o\B_o)}, $$
\e B= \frac{\%\B_t\.\%\A_t}{(\#u_t\.\%\A_t)(\#u_t\.\%\B_t)-\A_o\B_o} \f
$$  = \frac{\A_1\B_1+\A_2\B_2}{(\A_1\cos\VF+\A_2\sin\VF)(\B_1\cos\VF+\B_2\sin\VF)-\A_o\B_o}. $$

Because, for $\#k_t$ satisfying \r{De0}, any linear combination of incident and reflected plane waves satisfies the boundary conditions, the reflection cofficient can be considered as indeterminate. This phenomenon was first pointed out by Per-Simon Kildal as an anomaly arising in wave reflection from the then newly introduced DB boundary \cite{Kildal09,DB}. Later, a similar effect was shown to emerge when studying the SHDB boundary \cite{SHDB}. Actually, it does not appear obvious to picture a wave with normal incidence ($\#k_t=0$)  to the DB boundary as one similar to a surface wave.

Let us consider solutions of the dispersion equation \r{De0} for some special cases of the GSHDB boundary defined by real parameters $\A_i,\B_j$. Considering first the SHDB special case with $\A_o=\B_o$ and $\%\A_t=\%\B_t=\A_1\#e_1$, \r{De0} can be reduced to
\e (\A_1^2-\A_o^2)k_1^2 -(\A_ok_2^2-\A_1k_o)^2=0.\f
It is represented on the $\#k_t$ plane by a dispersion curve, which is an ellipse when $\A_1^2<\A_o^2$ and a hyperbola when $\A_1^2>\A_o^2$. For the special case of DB boundary with $\A_1=0$, the ellipse is reduced to the point $k_t=0$ while for the SH boundary with $\A_o=0$ the hyperbola is reduced to two parallel lines $k_t=\pm k_o/\cos\VF$. 

Assuming orthogonal vectors
\e \%\A_t=\A_1\#e_1,\ \ \ \ \%\B_t=\B_2\#e_2, \l{A1B2} \f
the GSHDB boundary does not reduce to a SHDB boundary. In this particular case we have $B=0$, whence one solution of \r{ktu} is $k_t=0$. The second solution yields
$$ \frac{1}{k_o}k_t(\VF) = \frac{\A_o\B_2\cos\VF - \B_o\A_1\sin\VF}{\A_1\B_2\cos\VF\sin\VF-\A_o\B_o} $$
\e = 1-\frac{(\A_1\sin\VF-\A_o)(\B_2\cos\VF + \B_o)}{\A_1\B_2\cos\VF\sin\VF-\A_o\B_o} .\l{ktVF} \f
At $\VF=\tan^{-1}(\A_o\B_2/\B_o\A_1)$ the normalized dispersion curve $k_t(\VF)/k_o$ passes through the origin  and it crosses the unit circle at $\VF=\sin^{-1}(\A_o/\A_1)$ if $|\A_o/\A_1|<1$ and $\VF=\cos^{-1}(-\B_o/\B_2)$ if $|\B_o/\B_2|<1$. When the normalized dispersion curve lies outside the unit circle, $k_3$ is imaginary and the matched plane wave equals a surface wave or a leaky wave.

In Figure 1, normalized dispersion curves are shown for four GSHDB boundaries in comparison with the unit circle, $\#k_t\.\#k_t=k_o^2$. The three concentric ellipses are defined by $\%\A_t=\#e_1$ and $\%\B_t=\#e_2$ and $\A_o=\B_o$ have the values 5, 2 and 1 corresponding to the respective smallest, middle-sized and largest ellipse. The fourth ellipse extending outside the unit circle is defined by $\A_o=\B_o=1$ and $\%\A_t=\#e_1\sqrt{2}$, $\%\B_t=\#e_2/\sqrt{2}$. For $\A_o=\B_o\ra\oo$ the dispersion curve approaches the point at the origin, corresponding to the DB boundary. In the fourth case, $\#k_t$ is complex for $\pi/2<\VF<3\pi/2$, whence the matched wave can be a surface wave. 

Figure 2 shows the locus of the unit vector $\#k/k_o$ of the matched plane wave for the same GSHDB boundary parameters as in Fig. 1. The curve outside the unit sphere is flat because the unit vector $\#k/k_o$ has an imaginary normal component.

\begin{figure}
	\centering
		\includegraphics[width=7cm]{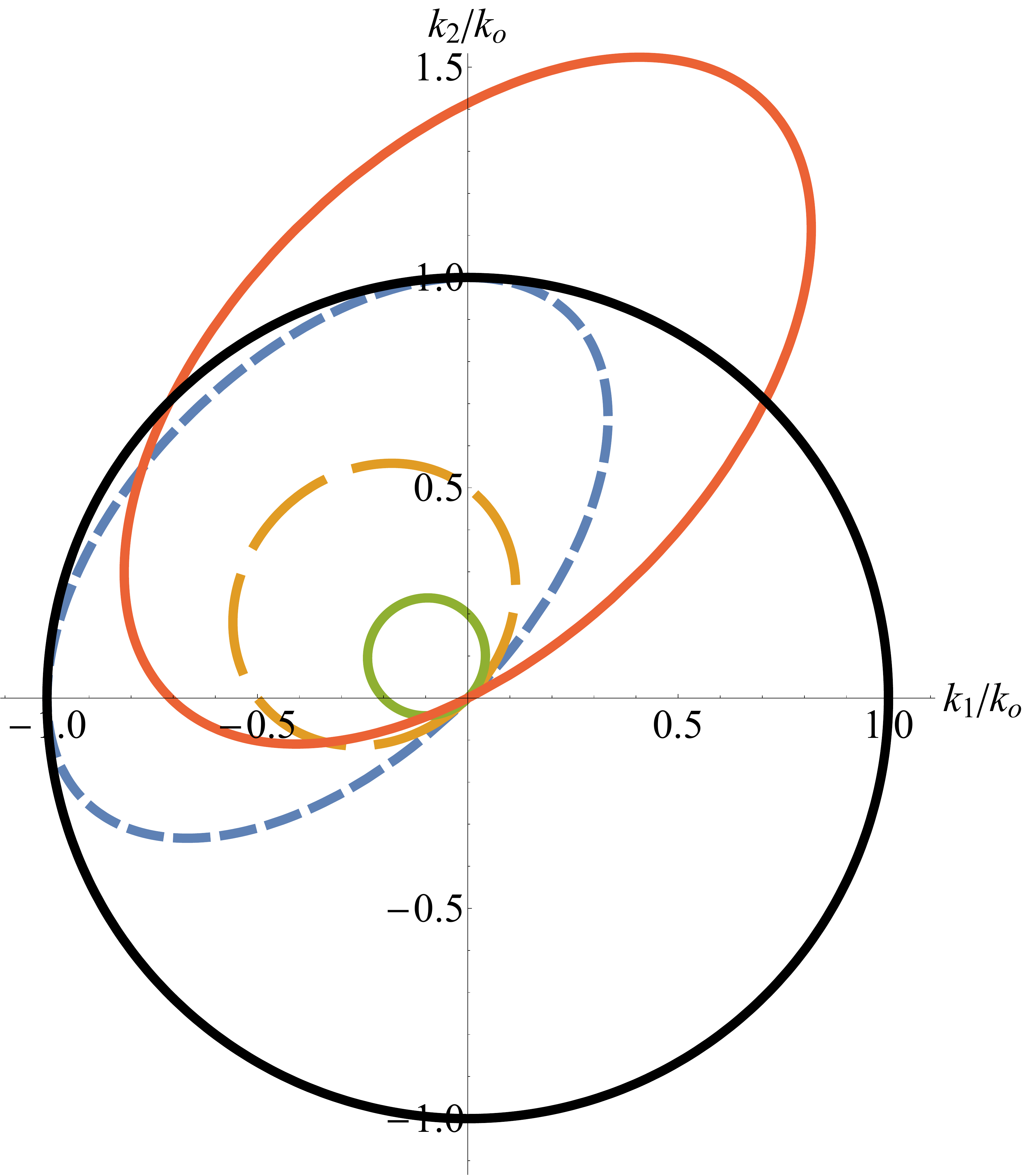}
	\label{fig:Fig1}
 \caption{Dispersion curves depicting $\#k_t/k_o$ for plane waves matched to a GSHDB boundary are defined by parameters $\%\A_t=\#e_1, \%\B_t=\#e_2$ and $\A_o=\B_o=5$ (solid), $2$ and $1$ (dashed) for the three ellipses within the unit circle in the order of growing in size. The ellipse extending outside the unit circle corresponds to $\%\A_t=\#e_1\sqrt{2}$ and $\%\B_t=\#e_2/\sqrt{2}$ , $\A_o=\B_o=1$}.
\end{figure}

\begin{figure}
	\centering
		\includegraphics[width=7cm]{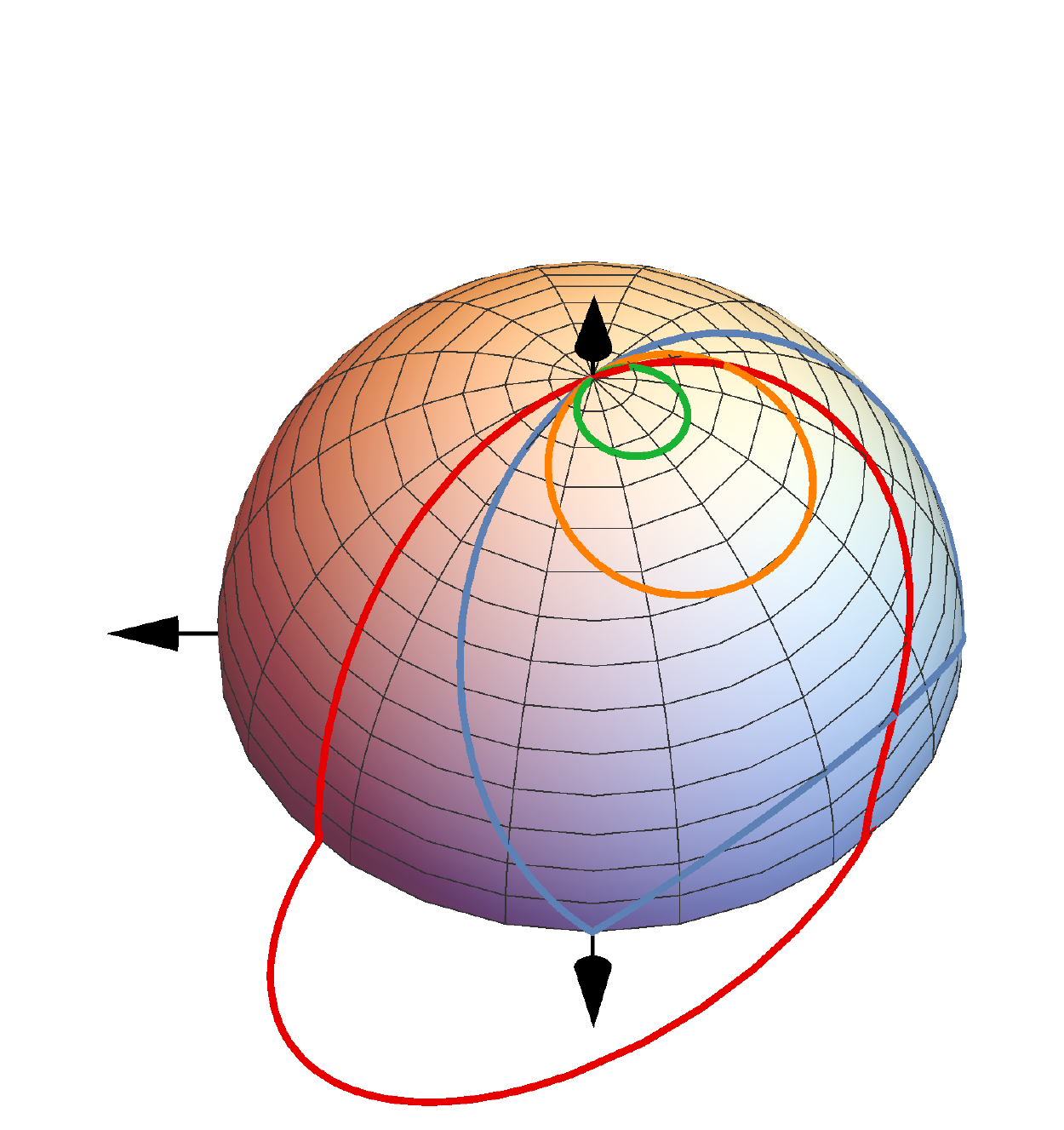}
	\label{fig:Fig2}
 \caption{Paths on the unit sphere corresponding to solutions of \r{De0} define directions of the normalized wave vector $\#k/k_o$ of a wave matched to the GSHDB boundary with the parameters of Figure1. The curve outside the sphere corresponds to the real part of the wave vector. The arrowheads define the unit vectors $\#e_3$ (vertical), $\#e_1$ and $\#e_2$.}
\end{figure}

\section{Medium interface as GSHDB boundary}

As a first step towards the realization of the GSHDB boundary with conditions \r{Bg} and \r{Dg}, let us consider an interface $\#e_3\.\#r=0$ of a bi-anisotropic medium defined by medium conditions of the form \cite{Methods}
\e \am \#D\\ \#B\a = \amm \=\E & \=\xi\\ \=\z & \=\M\a\.\am \#E\\ \#H\a. \f
To have correct boundary conditions at the interface, we can require the same conditions valid for the fields in the medium, 
\ea \B_4\#e_3\.(\=\E\.\#E+ \=\xi\.\#H) - \%\B_t\.\#H&=& 0 . \l{eq1}\\
\A_4\#e_3\.(\=\z\.\#E+ \=\M\.\#H) + \%\A_t\.\#E&=& 0, \l{eq2}\fa
Obviously, these will be satisfied for any fields when the medium dyadics are restricted by the conditions
\e \#e_3\.\=\E=0,\ \ \ \#e_3\.\=\xi=\%\B_t/\B_4,\f
\e \#e_3\.\=\z = -\%\A_t/\A_4, \ \ \ \#e_3\.\=\M=0, \f
which requires that the medium dyadics be of the form
\ea \amm \=\E & \=\xi\\ \=\z & \=\M\a &=&  \amm 0 & \#e_3\%\B_t/\B_4\\ -\#e_3\%\A_t/\A_4& 0\a \nonumber\\
&&+ \#e_3\x\amm \=\E' & \=\xi'\\ \=\z' & \=\M'\a. \l{medium}\fa
Here, $\=\E',\=\xi',\=\z'$ and $\=\M'$ may be any dyadics. Realization of such a medium by a metamaterial remains a challenge for the future.

As a simple special case, for $\%\A_t=\%\B_t=0$ we have the DB medium, which can be obtained at the interface an anisotropic medium satisfying $\#e_3\.\=\E=\#e_3\.\=\M=0$. This was suggested already in the article \cite{Rumsey} where the boundary conditions, later known as the DB conditions, were originally introduced. For the other simple special case, the SH boundary \cite{SHS1} with $\A_4=\B_4=0$ and $\%\A_t=\%\B_t=\#e_1$, \r{medium} does not apply. Starting from \r{eq1} and \r{eq2}, we see that the SH boundary conditions can be obtained at the interface of an anisotropic medium whose dyadic components approach infinite values as
\e |\=\E\.\#e_1|\ra\oo,\ \ \ \ |\=\M\.\#e_1|\ra\oo. \f
This corresponds to a medium consisting of parallel PEC and PMC planes orthogonal to $\#e_2$. Realization of the SH boundary by a tuned corrugated surface is a classic example of microwave engineering \cite{Cutler} and its extensions have been discussed by Per-Simon Kildal \cite{Kildal09}.

\section{Conclusion}

In the present paper, the previously introduced set of soft-and-hard/DB (SHDB) boundary conditions have been generalized one step further, and dubbed GSHDB conditions. The SHDB conditions were generalizations of the soft-and-hard (SH) and DB boundary conditions. It was shown how all of these boundary conditions can be naturally introduced applying four-dimensional formalism. In particular, the SHDB conditions depend on a single one-form while its generalization, the GSHDB conditions, are defined in terms of two one-forms. Plane wave reflection from a GSHDB boundary is analyzed in terms of conventional Gibbsian 3D vector and dyadic algebra. In particular, it is shown that any plane wave can be split in two eigenpolarizations, the e-wave and the m-wave, the former of which is reflected as from a PEC boundary, and the latter, as from a PMC boundary. The same property was previously shown to be valid for the SHDB boundary as well as for its two special cases, the SH boundary and the DB boundary. Particular attention is paid to the possibility of defining plane waves matched to a given GSHDB boundary, which contains surface waves and leaky waves as special cases. 2D dispersion equation for the plane wave matched to the GSHDB boundary, corresponding to the 3D dispersion equation for a plane wave in an electromagnetic medium, is derived and dispersion diagrams are depicted for some special cases. The possibility of realizing the GSHDB boundary as an interface of a bi-anisotropic medium was discussed. Being one of the basic concepts of electromagnetics due to its simple definition, the GSHDB boundary is of great theoretical interest. Since the SH and DB boundaries have found useful engineering applications and practical realizations as metasurface constructions, the GSHDB boundary can also be expected to have potential applications in the future.

\section{Acknowledgment} The authors wish to dedicate this paper to the memory of Per-Simon Kildal (1951-2016), who tragically passed away in April. Per-Simon was the originator of the concept of soft-and-hard (SH) boundary \cite{SHS1}, and the leading figure in studying its properties and applications over the years.

\end{document}